\documentclass[%
 aip,
 amsmath,amssymb,
 reprint,%
]{revtex4-1}
\usepackage{epsfig}
\usepackage{graphicx}
\usepackage{dcolumn}
\usepackage{bm}
\usepackage[T1]{fontenc}
\usepackage{mathptmx}
\usepackage{xcolor}
\usepackage{romannum}
\begin{document}
\preprint{AIP/123-QED}
\title[]{Multi-branched resonances, chaos through quasiperiodicity, 
 and asymmetric states in a superconducting dimer
}
\author{J. Shena}
\affiliation{National University of Science and Technology ``MISiS'', 
             Leninsky Prospekt 4, Moscow 119049, Russia}  
             \author{N. Lazarides}
\affiliation{Department of Physics, University of Crete, 71003 Herakleio, Greece}
\author{J. Hizanidis}
\email{hizanidis@physics.uoc.gr}
\affiliation{Department of Physics, University of Crete, 71003 Herakleio, Greece}

\date{\today}
\begin{abstract}
A system of two identical SQUIDs (superconducting quantum interference devices) 
symmetrically coupled through their mutual inductance and driven by a sinusoidal
field is investigated numerically with respect to dynamical properties such as 
its multibranched resonance curve, its bifurcation structure, as well as its 
synchronization behavior. The {\em SQUID dimer} is found to exhibit 
a hysteretic resonance curve with a bubble connected to it through Neimark-Sacker 
(torus) bifurcations, along with coexisting chaotic branches in their vicinity. 
Interestingly, the transition of the SQUID dimer to chaos occurs through a 
period-doubling cascade of a two-dimensional torus (quasiperiodicity-to-chaos 
transition). The chaotic states are identified through the calculated Lyapunov 
spectrum, and their basins of attraction have been determined. Bifurcation 
diagrams have been constructed on the parameter plane of the coupling strength
and the driving frequency of the applied field, and they are superposed to maps
of the maximum Lyapunov exponent on the same plane. In this way, a clear 
connection between chaotic behavior and torus bifurcations is revealed. Moreover, 
asymmetric states that resemble localized synchronization have been detected 
using the correlation function between the fluxes threading the loop of the 
SQUIDs. The effect of intermittent chaotic synchronization, which seems to be 
present in the SQUID dimer, is only slightly touched.     
\end{abstract}
\maketitle

{\bf Networks of coupled limit cycle oscillators represent a class of systems 
with special interest in physics, chemistry, and biological sciences. Arrays of
coupled nonlinear oscillators, in particular, which often exhibit complex 
dynamical behavior, have attracted large amounts of theoretical, numerical, and 
experimental efforts. However, the prominent features of a finite large number 
of such oscillators can be sometimes understood by analysing just two coupled
oscillators. A unique in many aspects oscillator is the superconducting quantum 
interference device (SQUID) that has been investigated extensively for many 
years. The SQUID is a low-loss, highly nonlinear resonant element that responds 
strongly to applied magnetic field(s) and exhibits rich dynamical behavior. 
Moreover, it has recently been the elementary unit for the construction of 
metamaterials in one and two dimensions. SQUIDs and SQUID arrays are 
technologically important devices, and they also serve as a testbed for 
exploring complex dynamics. Two SQUIDs in close proximity are coupled together 
through magnetic dipole-dipole forces, and their dynamical complexity significantly 
increases. Specifically, more complex bifurcation scenarios appear, along with 
transitions to chaos through quasiperiodicity, the emergence of localized 
synchronization, and intermittent chaotic synchronization. These dynamical 
properties are explored numerically with a well-established model whose 
parameters are acquired from recent relevant experiments.        
}

\section{\label{sec:level1} Introduction}
There has been great interest on the behavior of systems with interacting and
forced nonlinear oscillations with applications in physics, biology, and 
chemistry, which in addition are in abundance in the natural world 
\cite{Awrejcewicz1991}. The interaction introduces qualitatively new behavior 
to the network of oscillators, as compared to that of a single oscillator. For 
example, synchronization of the oscillating units of the system 
\cite{Rosenblum2003}, extreme multistability \cite{Wiesenfeld1989,Nishio1992}, 
multiple resonance and anti-resonance effects 
\cite{Jothimurugan2016,Kominis2019,Sarkar2019},
complex bifurcation structure \cite{Yin1998,Kenfack2003}, localization 
\cite{Lazarides2010}, emergence of exceptional points \cite{Kominis2018}, the 
intriguing effect of amplitude death \cite{Herrero2000}, chaos synchronization
\cite{Tafo2009}, 
localized synchronization \cite{Hohl1997}, chaos to hyperchaos transitions 
\cite{Kapitaniak1991,Kapitaniak1995,Kapitaniak2000,Grygiel2000}, and 
quasi-periodicity with subsequent transition to chaos 
\cite{Rand1982,Bishop1986,Kozlowski1995,Dixon1996,Elhadj2008,Borkowski2015}, 
are some of the experimentally or numerically observed behaviors.

Here we consider a pair of coupled  nonlinear oscillators (i.e., a nonlinear 
dimer) that belong to the large class of externally driven and dissipative 
systems. More specifically, we consider a pair of identical mutually coupled 
SQUID oscillators \cite{Kleiner2004,Fagaly2006}, where the acronym stands for 
``superconducting quantum interference device''. SQUIDs are mesoscopic 
superconducting devices which are modeled efficiently by equivalent electrical 
circuits, while their dynamics is governed by a second order nonlinear ordinary 
differential equation. The simplest variant of a SQUID consists of a 
superconducting loop interrupted by a Josephson junction (JJ), called rf SQUID,
which was found to exhibit very rich dynamic behavior, including complex 
bifurcation structure and chaos \cite{Hizanidis2018}. Moreover, rf SQUIDs are
employed as elementary units for the fabrication of superconducting metamaterials,
whose investigation has revealed several extraordinary properties 
\cite{Trepanier2013,Zhang2015}.   
Recent theoretical works on SQUID metamaterials have also reported the existence of chimera states and pattern formation~\cite{LAZ15,HIZ16a,HIZ16b,HIZ20}.

\begin{figure}[!h]
   \includegraphics[scale=0.2]{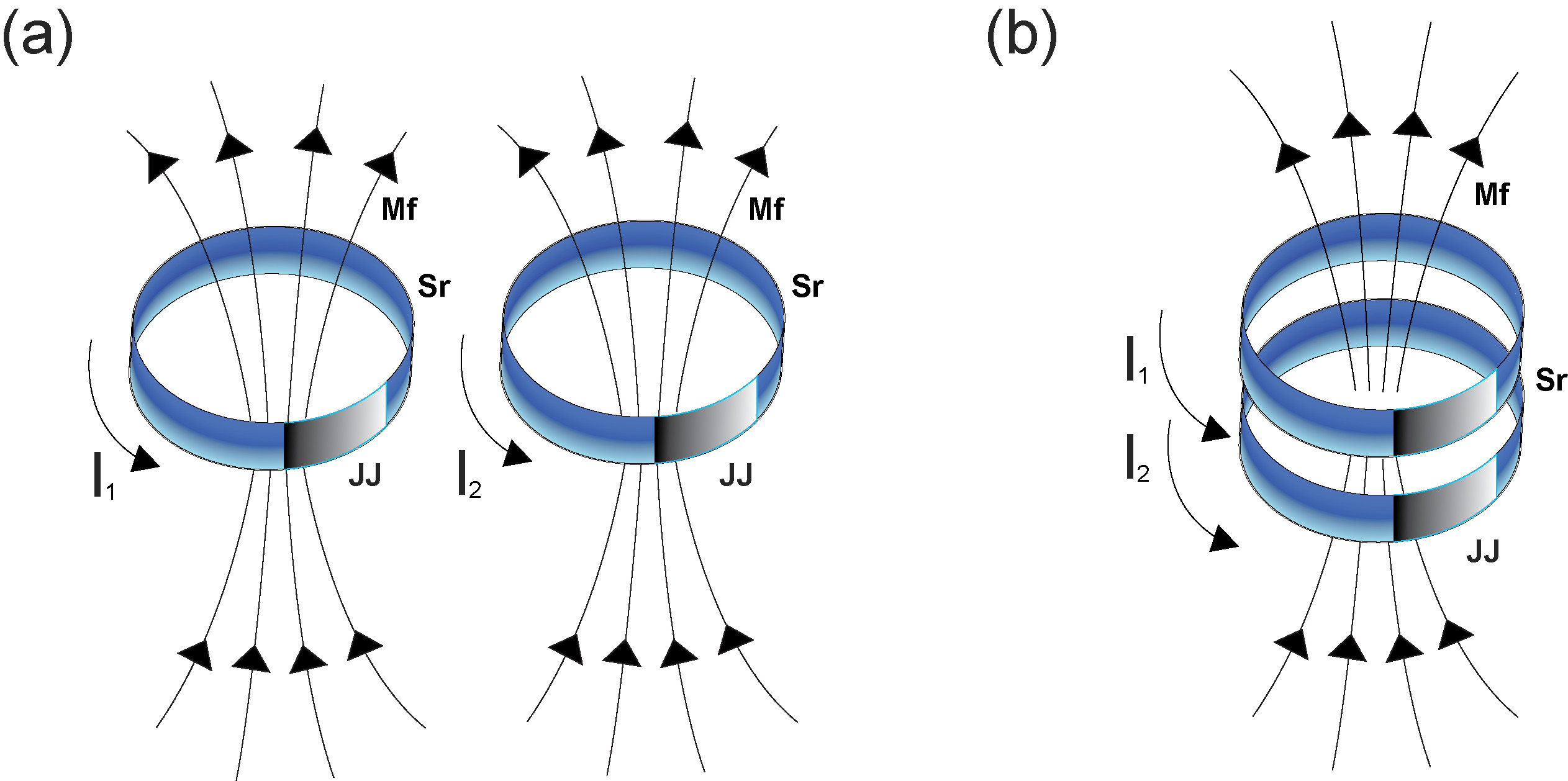}
   \caption{Schematic diagram of a SQUID dimer in a magnetic field where ``Mf'' 
            is the Magnetic field, ``Sr'' is the Superconducting ring, ``JJ'' is 
            the Josephson Junction, and $I_1$, $I_2$ are the induced currents. 
            Panels (a) and (b) correspond to negative and positive magnetic
            coupling strength, respectively, appropriate for the planar and the 
            axial geometry.}
\label{fig1}
\end{figure}
Two SQUIDs in relatively close proximity are coupled together by magnetic 
dipole-dipole forces due to their mutual inductance $\cal M$, while the sign and 
the magnitude of their coupling strength coefficient depends strongly on their 
relative positions. Specifically, the two SQUIDs may be arranged either in the 
planar geometry (Fig. \ref{fig1}(a)), where they lie in the same plane with 
their axes in parallel, or in the axial geometry, they lie the one on top of the
other with their axes lying on the same line (Fig. \ref{fig1}(b)). The SQUIDs are
driven by externally applied periodic and constant (bias) magnetic fields of 
appropriate polarization, to which they respond resonantly. Although its 
potential for technological applications, the dynamics of a SQUID dimer are not 
properly investigated. Recently however, the existence of homoclinic chaos in a
pair of parametrically driven SQUIDs has been exposed 
\cite{Agaoglou2015,Agaoglou2017}.

In the present work, particular dynamical properties of a SQUID dimer are 
addressed, such as its highly unusual resonance curve that consists of a 
bistable part with a bubble connected to it through Neimark-Sacker (NS) and 
pitchfork bifurcations (PBs). These bubbles are of very different character compared to those reported 
long ago by Bier and Bountis~\cite{BIE84} and they deserve to 
be investigated on their own. Moreover, in addition to those solutions, there 
are at least two more coexisting solutions in the frequency interval spanned by 
the resonance curve, which end up becoming chaotic through quasiperiodicity, 
i.e., through period doubling bifurcations of two-dimensional (2D) tori. The 
basins of attraction leading to chaos are calculated for two parameter sets of 
particular importance. Furthermore, asymmetric solutions either synchronized 
(leading to localized synchronization) or anti-synchronized (i.e., having 
opposite phases) are investigated by calculating the maximum Lyapunov exponent 
and the correlation function over the entire control parameter space.

In the next section (II), the electrical circuit equivalent model for the SQUID 
dimer is described in detail, and the dynamic equations are obtained. In section
III, the bifurcation structure for the SQUID dimer is discussed along with the 
emergence of quasiperiodicity and chaos. Section IV is devoted to the transition
to chaos through quasiperiodicity, through a Neimark-Sacker (NS) bifurcation. The 
existence of highly asymmetric solutions related to localized synchronization is 
revealed using an appropriate correlation function in section V. 
Conclusions are given in section VI.

\section{Modeling Two Coupled SQUIDs}
\label{sec:model}
Consider two identical rf SQUIDs in close proximity so that they are coupled by 
magnetic dipole-dipole forces through their mutual inductance $\cal M$ (Fig. 
\ref{fig1}). Each of the SQUIDs is modeled by an equivalent electrical circuit 
that features a self-inductance $L$ due to the superconducting ring, which is 
connected in series with a ``real'' Josephson junction \cite{Josephson1962} 
characterized by a critical current $I_c$, capacitance $C$, and Ohmic resistance 
$R$ \cite{Hizanidis2018}. The SQUIDs are subject to an externally applied 
spatially constant and time-periodic (ac) magnetic field and a constant in time 
and space (dc) magnetic field. Those fields add an electromotive force in series 
to the SQUID's equivalent circuits. The magnetic flux threading the loops of the 
SQUIDs includes supercurrents as well as normal (i.e., quasiparticle) currents 
around the SQUID's rings through Faraday's law. In turn, the induced currents 
produce their own magnetics field which counter-acts the applied ones. Then, the 
flux $\Phi_1$ and $\Phi_2$ through the loop of the SQUID number $1$ and $2$, 
respectively, is given by the following {\em flux-balance relations}         
\begin{eqnarray}
\label{2.01}
   \Phi_1 =\Phi_{ext} +L\, I_1 +{\cal M} I_2, 
\\
\label{2.02}
   \Phi_2 =\Phi_{ext} +L\, I_2 +{\cal M} I_1,
\end{eqnarray}
where $I_1$ and $I_2$ are the currents induced by the external magnetic fields,
and the external flux 
\begin{equation}
\label{2.03}
   \Phi_{ext} =\Phi_{dc} +\Phi_{ac}\, \cos(\omega t),
\end{equation} 
with $\Phi_{dc}$ the dc flux bias, $\Phi_{ac}$ the amplitude of the ac flux,
$\omega$ its frequency, and $t$ the temporal variable. Eqs. (\ref{2.01}) and
(\ref{2.02}) can be solved for the currents and written in matrix form as  
\begin{eqnarray} 
\label{2.04}
  \begin{bmatrix}
    I_1 \\
    I_2 
  \end{bmatrix}
  =\frac{1}{L}
\begin{bmatrix}
    1       & \lambda \\
    \lambda & 1 
  \end{bmatrix}^{-1}
  \begin{bmatrix}
    \Phi_1 -\Phi_{ext} \\
    \Phi_2 -\Phi_{ext}
  \end{bmatrix},
\end{eqnarray} 
where $\lambda=\frac{\cal M}{L}$ is the magnetic coupling strength between the 
SQUIDs. Note that the sign of $\lambda$ depends on the mutual position of the two
SQUIDs. In the planar geometry (Fig. \ref{fig1}(a)), that sign is negative, 
while in the axial geometry (Fig. \ref{fig1}(b)), it is positive. The currents 
flowing in the SQUIDs are given within the framework of the resistively and 
capacitively shunted junction (RCSJ) model \cite{Likharev1986}, as 
\begin{equation}
\label{2.05}
    I_n =-C \frac{d^2 \Phi_n}{dt^2} -\frac{1}{R} \frac{d \Phi_n}{dt} 
     -I_c\, \sin\left( 2 \pi \frac{\Phi_n}{\Phi_0} \right), 
\end{equation} 
where $\Phi_0$ is the flux quantum, and $n=1,2$. Combining Eqs. (\ref{2.04}) and
(\ref{2.05}) we get the coupled equations for the fluxes through the loops of 
the SQUIDs in natural units, as
\begin{eqnarray}
\label{2.06}
   L\, C \frac{d^2 \Phi_1}{dt^2} +\frac{L}{R} \frac{d \Phi_1}{dt}
   +L\, I_c\, \sin\left( 2 \pi \frac{\Phi_1}{\Phi_0} \right) 
   +\frac{1}{1-\lambda^2} \Phi_1 
\nonumber \\
   -\frac{\lambda}{1-\lambda^2} \Phi_2 =\frac{1}{1+\lambda} \Phi_{ext}, \\
\label{2.07}
   L\, C \frac{d^2 \Phi_2}{dt^2} +\frac{L}{R} \frac{d \Phi_2}{dt}
   +L\, I_c\, \sin\left( 2 \pi \frac{\Phi_2}{\Phi_0} \right) 
   +\frac{1}{1-\lambda^2} \Phi_2 
\nonumber \\
   -\frac{\lambda}{1-\lambda^2} \Phi_2 =\frac{1}{1+\lambda} \Phi_{ext},
\end{eqnarray}
In normalized form, Eqs. (\ref{2.06}) and (\ref{2.07}) can be written as
\begin{eqnarray}
\label{2.08}
   \ddot{\phi_1} +\gamma \dot{\phi_1}
   +\beta\, \sin\left( 2 \pi \phi_1 \right) 
   +\frac{1}{1-\lambda^2} \phi_1 
   -\frac{\lambda}{1-\lambda^2} \phi_2
\nonumber \\
   =\frac{1}{1+\lambda} \phi_{ext},
\\
\label{2.09}
   \ddot{\phi_2} +\gamma \dot{\phi_2}
   +\beta\, \sin\left( 2 \pi \phi_2 \right) 
   +\frac{1}{1-\lambda^2} \phi_2 
   -\frac{\lambda}{1-\lambda^2} \phi_1
\nonumber \\
   =\frac{1}{1+\lambda} \phi_{ext},
\end{eqnarray}
with
\begin{equation}
\label{2.10}
   \phi_{ext} =\phi_{dc} +\phi_{ac}\, \cos\left( \Omega \tau \right),
\end{equation}
where all the fluxes $\Phi_1, \Phi_2, \Phi_{ac}, \Phi_{dc}$ are in units of 
the flux quantum $\Phi_0$ and the new temporal variable $\tau$ is defined as 
$\tau =t\, \omega_{LC}$ with $\omega_{LC} =1/\sqrt{L\, C}$ being the
inductive-capacitive ($L\, C$) SQUID frequency. Note that the normalized driving 
frequency $\Omega$ is in units of $\omega_{LC}$, i.e., 
$\Omega =\omega/\omega_{LC}$. The overdots in Eqs. (\ref{2.08}) and (\ref{2.09}) 
denote differentiation with respect to the normalized temporal variable $\tau$. 
The rescaled SQUID parameter and the loss coefficient are given respectively by  
\begin{equation}
\label{2.11}
   \beta =\frac{L\, I_c}{\Phi_0}, \qquad \gamma =\frac{1}{R} \sqrt{ \frac{L}{C} }.
\end{equation} 
\begin{figure}[!t]
   \includegraphics[scale=0.22]{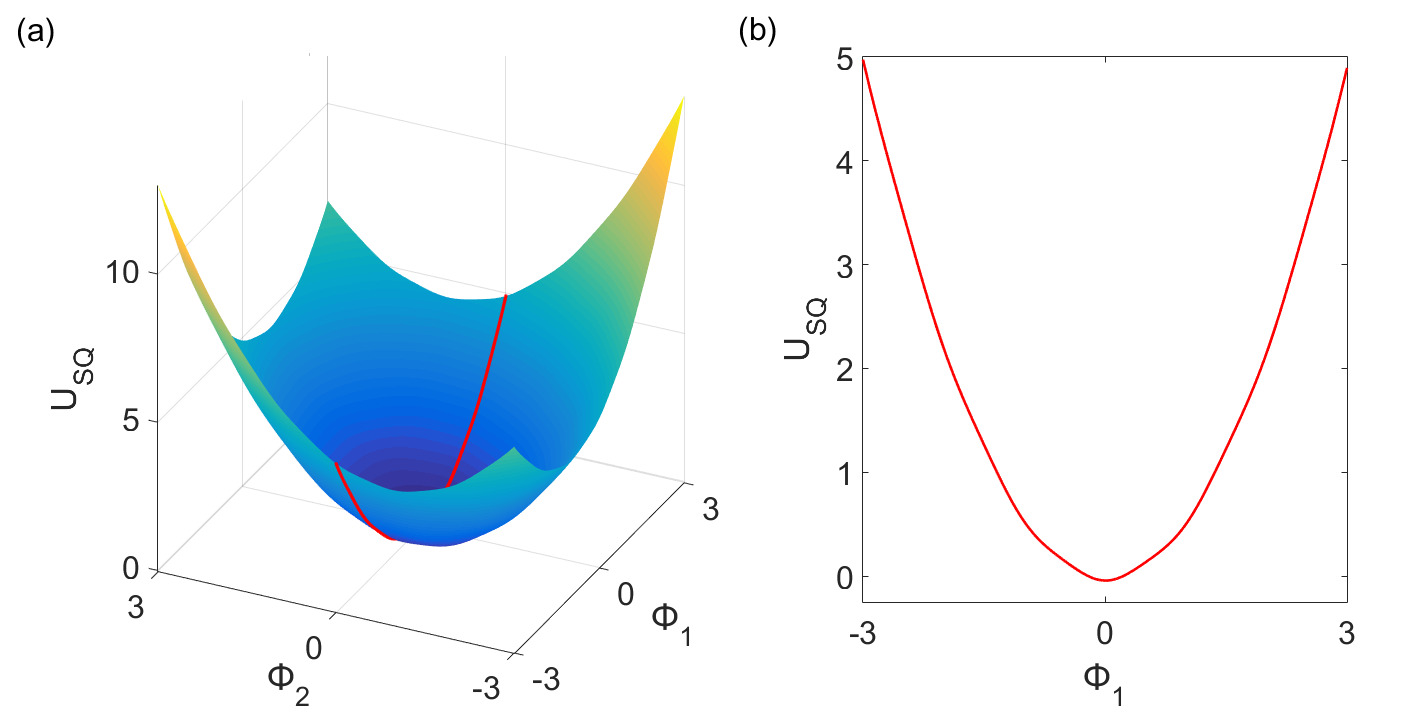}
   \caption{(a) The potential $U_{SQ}(\phi_1,\phi_2)$ of the SQUID dimer for 
            $\tau =T =2\pi/\Omega$ where $\Omega =0.891$. 
            (b) The corresponding one-dimensional projection of 
            $U_{SQ}(\phi_1,\phi_2)$ on the $\phi_2 =0$ plane. 
            Other parameters: $\lambda=0.31$, $\phi_{ac}=0.02$, $\gamma=0.024$, 
            and $\beta=0.1369$.}
\label{fig2}
\end{figure}
In the lossless case, i.e., when $\gamma=0$ ($R \rightarrow \infty$), the 
(Newtonian) dynamics of the SQUID dimer is governed by the equations
\begin{equation}
\label{2.12}
   \ddot{\phi}_1 =-\frac{\partial U_{SQ} (\phi_1, \phi_2)}{\partial \phi_1},
\qquad
   \ddot{\phi}_2 =-\frac{\partial U_{SQ} (\phi_1, \phi_2)}{\partial \phi_2} 
\end{equation}
in terms of the SQUID dimer potential
\begin{eqnarray}
\label{2.13}
   U_{SQ} (\phi_1,\phi_2) &=&-\frac{1}{1 +\lambda} 
                         \left( \phi_1 +\phi_2 \right) \phi_{ext}(\tau)
\nonumber \\
                        & &-\frac{\beta}{2\pi} 
        \left[ \cos( 2 \pi \phi_1 ) +\cos( 2 \pi \phi_2 ) \right]  
\nonumber \\
                        & &+\frac{1}{2 ( 1 - \lambda^2 )} 
        \left( \phi_1^2 -2\lambda \phi_1 \phi_2 +\phi_2^2 \right).
\end{eqnarray}
\begin{figure*}[!t]
   \includegraphics[scale=0.38]{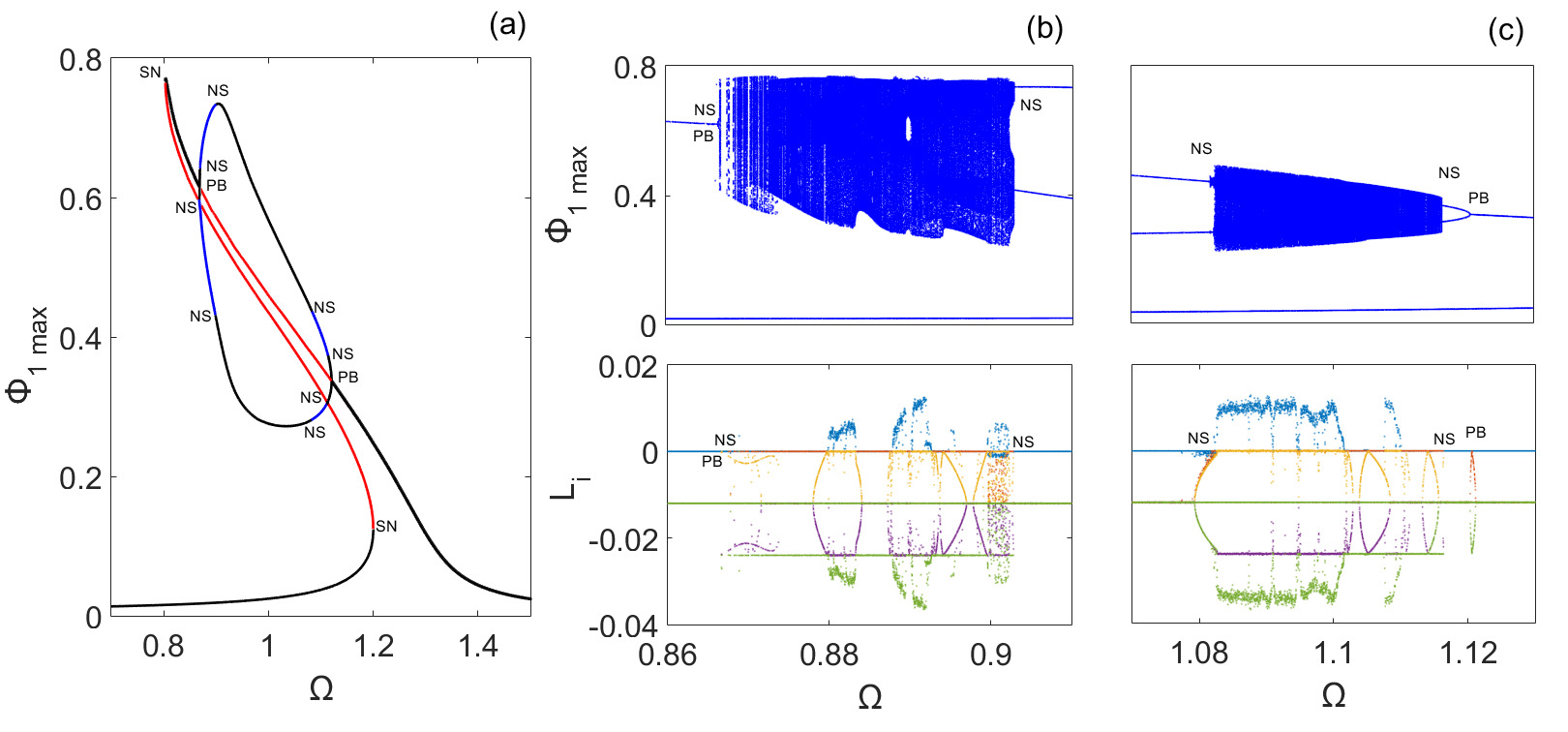}
   \caption{(a) The resonance curve of the SQUID dimer. Black lines correspond 
            to stable branches and red and blue lines correspond to unstable 
            ones. The latter are born through saddle-node bifurcations of limit 
            cycles (SN), Neimark-Sacker bifurcations (NS), or pitchfork 
            bifurcations of limit cycles (PB). 
            (b) Maximum values of the magnetic flux $\phi_1$ (top) and all five 
            Lyapunov exponents (bottom), as a function of the driving frequency 
            in the interval corresponding to the left blue branches of (a). 
            For each value of $\Omega$, $n_i =50$ different initial conditions
            were used. 
            (c) The same as in (b) for the right blue branches of (a). 
            Other parameters: $\lambda=0.31$, $\phi_{ac}=0.02$, $\gamma=0.024$, 
            and $\beta=0.1369$.}
\label{fig3}
\end{figure*}
Since the SQUID dimer considered here consists of two identical non-hysteretic 
SQUIDs ($2\pi\beta<1$), the potential $U_{SQ} (\phi_{1},\phi_{2})$ is a 
two-dimensional corrugated parabola with a single minimum that depends 
periodically on time $\tau$. A snapshot of the potential at $\tau =T$, with 
$T =2\pi/\Omega$ being the period of the (driving) external periodic field, is 
shown in Fig.~\ref{fig2}(a) while its one-dimensional projection on the 
$\phi_2 =0$ plane is shown in Fig.~\ref{fig2}(b).

In what follows, the external dc flux is set to zero, i.e., $\phi_{dc} =0$. 
For obtaining the values of the SQUID-dependent normalized parameters $\beta$ 
and $\gamma$ that go into the model equations (\ref{2.08}) and (\ref{2.09}), we 
choose $L=120 ~pH$, $C=1.1 ~pF$, $R=500 ~\Omega$, and $I_c =2.35 ~\mu A$ for 
their equivalent lumped-circuit elements. These values are typical for 
non-hysteretic SQUIDs investigated recently in the context of SQUID 
metamaterials \cite{Trepanier2013,Zhang2015}. By substituting them into  Eq. 
(\ref{2.11}), we get $\beta_L =0.86$ ($\beta \simeq 0.1369$) and $\gamma =0.024$,
while the $L C$ or {\em geometrical} frequency is 
$f_{LC} =\omega_{LC} /(2\pi) \simeq 13.9 ~GHz$. As mentioned earlier, the value 
of the coupling strength $\lambda$  depends significantly on the relative 
positions of the two SQUIDs, and may assume its value from the relatively broad
range of $[-0.06, +0.50]$ that corresponds to technologically feasible SQUID 
dimer designs \cite{Anlage0000}. Besides, there are also the parameters 
$\phi_{ac}$ and $\Omega$ ($\phi_{dc} =0$), which can be tuned externally. Here,
however, we prefer to fix the former to $\phi_{ac} =0.02$, which is well into 
the range of the experimentally accessible values, and use $\lambda$ and $\Omega$
as control parameters.

\section{Bifurcations, Quasi-periodicity, and Chaos}
It is known that for relatively high $\phi_{ac}$, the single SQUID exhibits a
multistable response which is reflected in its corresponding ``snake-like'' 
resonance curve~\cite{Hizanidis2018}, i.e., the curve formed by the maximum flux 
within one driving period $T=2\pi/\Omega$, $\phi_{max}$, as a function of the 
driving frequency $\Omega$. For the value of $\phi_{ac}=0.02$ considered here, 
the single SQUID is bistable. When two SQUIDs are coupled positively (i.e., when 
$\lambda > 0$) the bistability is maintained but the dynamics presents additional 
complexity. The resonance curve for the SQUID dimer is plotted in 
Fig.~\ref{fig3}(a) in terms of the magnetic flux $\phi_1$ threading the loop of 
the SQUID number $1$. Specifically, the maximum flux threading the loop of the 
SQUID number $1$ within one driving period $T$, $\phi_{1,max}$, is plotted 
against the driving frequency $\Omega$. 
We see the typical tilted curve associated with bistability and hysteresis, 
where two stable solutions (black curves) coexist with an unstable one (red 
curve) below the geometric resonance frequency $\Omega =1$. Starting at lower driving frequencies and following the periodic 
solution in $\Omega$ there is a saddle-node bifurcation of limit cycles (SN) 
occurring at $\Omega=1.2$. The solution then becomes unstable and when it 
reaches the peak of the resonance curve it turns stable again in a second SN 
bifurcation at $\Omega=0.8036$. The stable branch that is born then undergoes a 
pitchfork bifurcation of limit cycles (PB) at $\Omega=0.8684$, where in addition 
to the unstable branch two more stable branches of periodic solutions are born. 
These branches consecutively become unstable in a Neimark-Sacker bifurcation (NS) 
on either side at $\Omega=0.87$, shown in blue color. The blue branches become stable again in a 
second NS bifurcation at $\Omega=0.8944$, again on either side. The same scenario takes place at 
$\Omega=1.12$ (right PB) and this family of solutions emanating from the two 
``opposite to each other'' PB bifurcations, undergoes NS bifurcations (at $\Omega=1.083$ and $\Omega=1.114$) on 
either side forming, thus, a closed loop. 
The bifurcation lines have been obtained 
using a very powerful software tool that executes a root-finding algorithm for 
continuation of periodic solutions~\cite{ENG02}.

We focus now on the two $\Omega$ intervals corresponding to the blue unstable 
branches of Fig.~\ref{fig3}(a) which lie within the two PB bifurcations on both right and left side. 
In the upper panels of Figs.~\ref{fig3}(b) 
and (c) the maxima of the solution for $\phi_1 (\tau)$ are plotted over $\Omega$ 
for the left and right interval, respectively. Note that at each frequency 
$\Omega$, a large number of different initial conditions $n_i$ were used (here 
$n_i =50$) in order to obtain all the coexisting solutions within the $\Omega$ 
interval shown, and also to obtain a good representation of possible chaotic 
behavior. Apparently, chaotic behavior appears in both the upper panels of 
Figs.~\ref{fig3}(b) and (c). These solution branches coexist with the blue 
unstable branches of Fig.~\ref{fig3}(a) mentioned above; the reason for which 
they are not superposed to them is merely clarity.

In the corresponding lower 
panels, the Lyapunov exponents of the obtained solutions that are plotted for 
the same intervals of $\Omega$, reveal chaotic behavior in the regions where 
the maximum Lyapunov exponent $L_1$ is positive (positive Lyapunov exponents are 
shown as blue points). Careful inspection of the Lyapunov exponents indicate 
that transitions to chaos through quasiperiodicity and reversely take place in 
Figs.~\ref{fig3}(b) and (c) (see next section). The Lyapunov exponents were calculated employing the algorithm from \cite{Geist1990} using 
the Julia$^R$ software library. Similar multibranched resonance curve are 
sometimes encountered in dissipative-driven oscillators 
\cite{Warminski2015,Marchionne2018,Zang2019} where in addition, isolated 
branches (``isolas'') may exist. However, to the best of our knowledge, 
multibranched resonance curves whose branches are connected through 
Neimark-Sacker bifurcations have never before been observed.

\begin{figure}
   \includegraphics[scale=0.4]{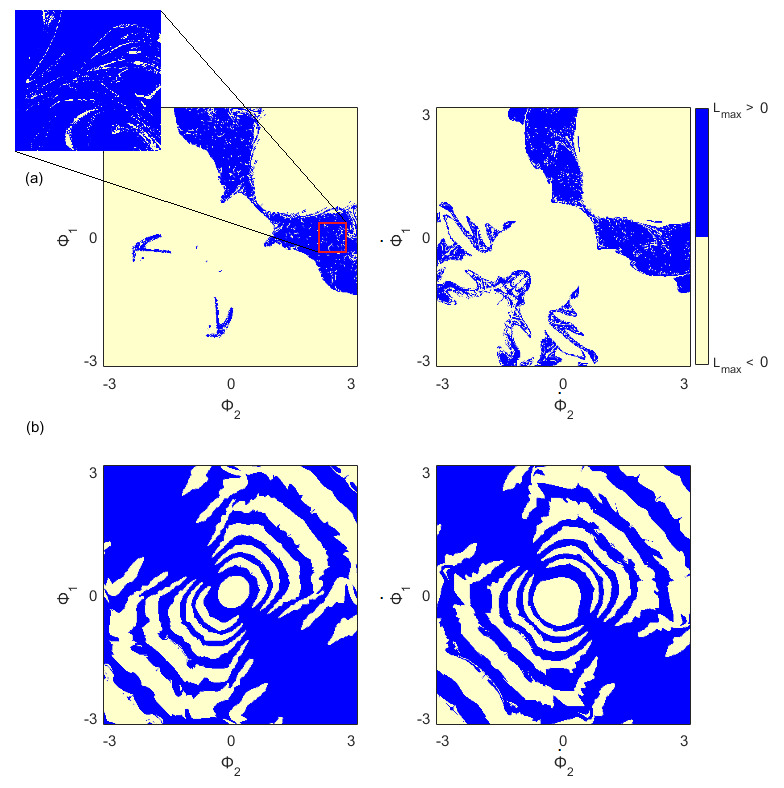}
   \caption{Basins of attraction for the SQUID dimer in terms of the maximum 
            Lyapunov exponent $L_{\text{max}}$. Blue and yellow color 
            corresponds to chaotic and non-chaotic dynamics, respectively.
	    Left: $\phi_1$ versus $\phi_2$ for 
            $(\dot{\phi}_1(0),\dot{\phi_2}(0)) = (0,0)$. 
            Right: $\dot{\phi}_1$ versus $\dot{\phi}_2$ for 
            $(\phi_1(0),\phi_2(0)) = (0,0)$. 
            (a) $\Omega=0.891$, and (b) $\Omega=1.1$. The inset in the left panel of (a)
            shows a blow-up of the region inside the red box. Other parameters: 
            $\lambda=0.31$, $\phi_{ac}=0.02$, $\gamma=0.024$, and $\beta=0.1369$.} 
\label{fig4}
\end{figure}

Since the system dynamics is very sensitive to initial conditions, we have 
identified the basins of attraction of chaotic and periodic behavior for two 
$\Omega$ values within the intervals shown in Figs.~\ref{fig3}(b) and (c) 
close to the NS bifurcations of limit cycles. 
These are displayed in Figs.~\ref{fig4}(a) and (b), respectively, both in the 
$(\phi_1,\phi_2)$ plane (left panels) and $(\dot{\phi}_1, \dot{\phi}_2)$ plane 
(right panels). Blue (dark) regions denote the set of initial conditions leading 
to chaotic motion, while light yellow (light) regions the ones leading to 
periodic or quasiperiodic oscillations. The other state variables are initially 
fixed as $\dot{\phi}_1 (\tau=0) =\dot{\phi}_1 (\tau=0) =0$ and 
${\phi}_1 (\tau=0) ={\phi}_1 (\tau=0) =0$, respectively. In Fig.~\ref{fig4}(a),
an interesting intertwining between the basins of attraction of chaotic and 
periodic or quasiperiodic states can be seen. This is apparent from the presence
of yellow spots located erratically within the blue regions in both subfigures, 
that resemble riddled basins investigated in various low-dimensional systems
\cite{Lai1994,Sathiyadevi2019}. The inset of Fig.~\ref{fig4}(a) reveals this intricate structure of the basin of attraction on a smaller scale. Note that a classification of basins of 
attraction in several low-dimensional systems has been perormed in Ref. 
\cite{Sprott2015}. The knowledge of the basins is essential for determining the 
usefulness of the systems in practical applications.

The other control parameter in our system is the coupling strength $\lambda$, 
which (as described in Section~\ref{sec:model}) can obtain both negative and 
positive values within a certain range provided by calculations on feasible 
SQUID dimer designs \cite{Anlage0000}. In Fig.~\ref{fig6}(a) the co-dimension 2 
bifurcation diagram is shown in the $(\Omega,\lambda)$ parameter plane. Black 
lines mark the continuation of the saddle-node bifurcations and red lines that 
of the Neimark-Sacker (torus) bifurcations. In Fig.~\ref{fig6}(b) the maximum Lyapunov 
exponent is plotted in the $(\Omega,\lambda)$ parameter space. Blue and yellow 
regions correspond to chaotic (positive $L_{\text{max}}$) and non-chaotic 
(negative or zero $L_{\text{max}}$) dynamics, respectively. We can see that the 
chaotic regimes lie within the boundaries of the NS bifurcation curves. In the 
following section we explore in detail the transition from quasiperiodicity to 
chaos in our system.

\section{Torus-doubling transition to chaos}
\begin{figure}[!t]
   \includegraphics[scale=0.22]{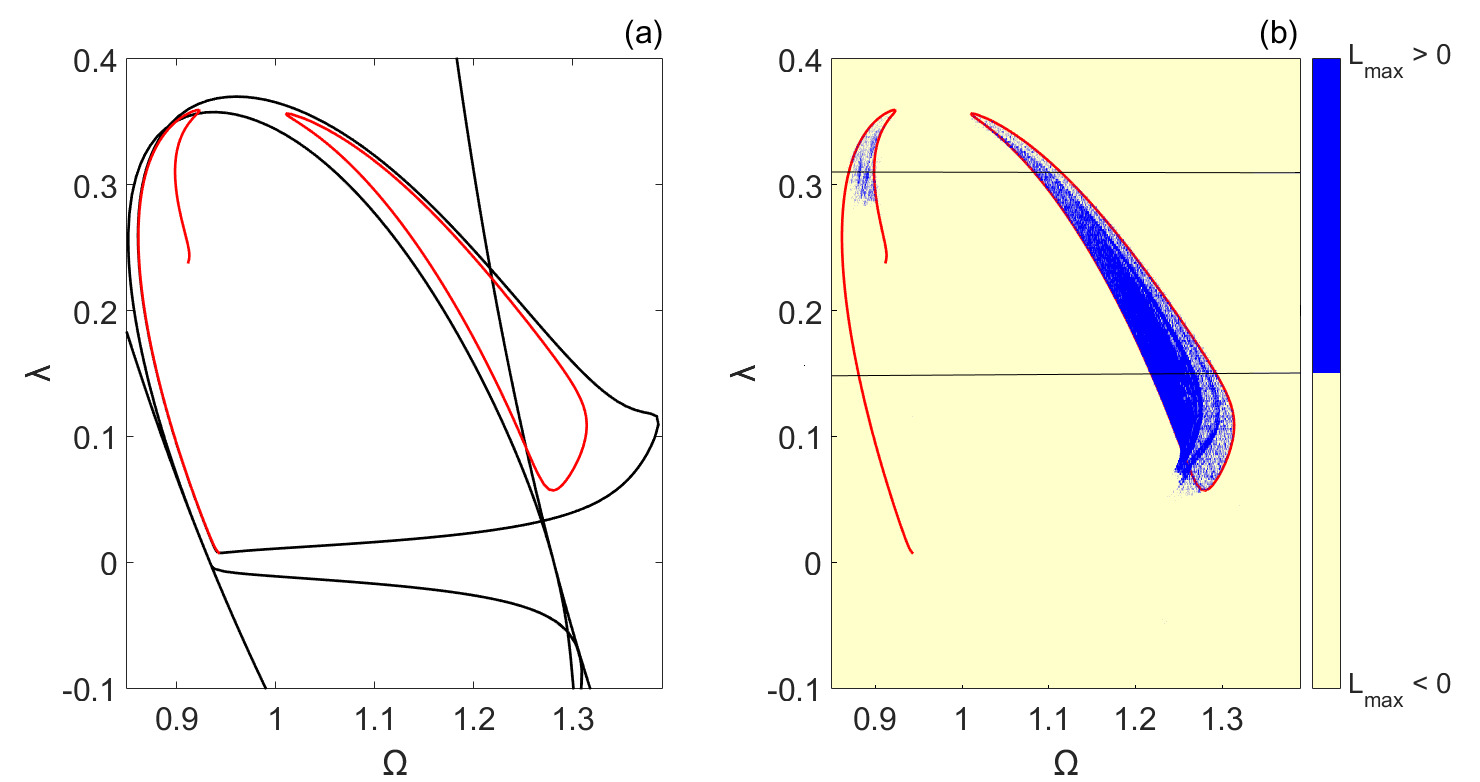}
   \caption{(a) Bifurcation diagram in the ($\Omega,\lambda$) parameter plane of 
            the two coupled SQUIDs. Black and red lines correspond to saddle-node 
            (SN) bifurcations of limit cycles and Neimark-Sacker (NS) 
            bifurcations, respectively. 
            (b) Map of the maximum Lyapunov exponent $L_{\text{max}}$ in the 
            ($\Omega, \lambda$) parameter space, on which the NS bifurcation 
            curve from (a) is superposed (red lines). Blue and yellow color 
            corresponds to chaotic and non-chaotic dynamics, respectively.  
            Other parameters:  
            $\phi_{ac}=0.02$, $\gamma=0.024$, and $\beta=0.1369$.}
\label{fig5}
\end{figure}

From Figs.~\ref{fig3}(b)-(c) and ~\ref{fig5}(b) it is clear that the SQUID dimer 
described by Eqs. (\ref{2.08}) and (\ref{2.09}) becomes chaotic through
Neimark-Sacker bifurcations. This quasiperiodic transition 
to chaos is a well-known scenario whereby a torus in low-dimensional dynamical 
systems loses its stability and develops into chaos. A path from a limit cycle, 
then a transition into 2D torus following by a 3D torus and eventually direct 
transition to chaos has been presented numerically and experimentally is an 
analog electrical circuit representing the ring of unidirectionally coupled 
single-well Duffing oscillators \cite{Borkowski2015}. Another possible scenario 
to chaos is a cascade of period doublings of the torus which has been 
investigated numerically long ago using low-dimensional maps 
\cite{Arneodo1983,Kaneko1984}, and experimentally in electrochemical reactions
\cite{Bassett1989}. According to the second scenario, a finite cascade of period 
doubling of such invariant tori leads eventually to chaos. This scenario has 
been reported numerically for a quintic complex Ginzburg-Landau equation 
\cite{Kim1997} and bimodal laser model \cite{Letellier2007} as well as 
experimentally, in Rayleigh-Benard convection \cite{Flesselles1994}, in a simple
thermoacoustic system \cite{Mondal2017}, and near the ferroelectric phase 
transition of $KH_{2}PO_{4}$ crystals \cite{Shin1999}.

\begin{figure}[!b]
   \includegraphics[scale=0.24]{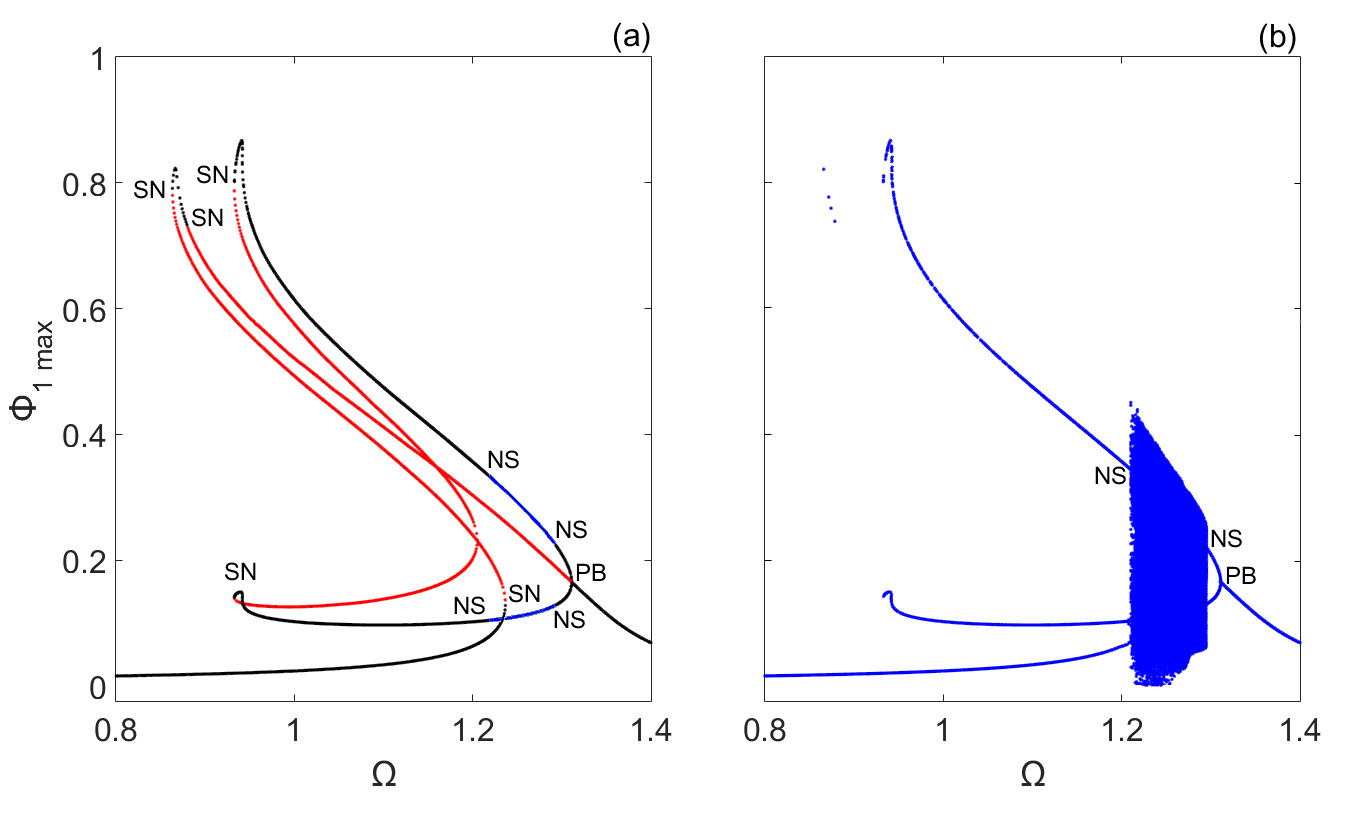}
   \caption{(a) The resonance curve of the SQUID dimer for $\lambda=0.15$. Black lines correspond 
            to stable branches and red and blue lines correspond to unstable 
            ones. The latter are born through saddle-node bifurcations of limit 
            cycles (SN), Neimark-Sacker bifurcations (NS), or pitchfork 
            bifurcations of limit cycles (PB). 
            (b) Maximum values of the magnetic flux $\phi_1$ of stable solutions as a function of the driving frequency 
            in the same $\Omega$ interval.
            Other parameters: $\phi_{ac}=0.02$, $\gamma=0.024$, 
            and $\beta=0.1369$.
            }
\label{fig6}
\end{figure}

 \begin{figure}[!b]
   \includegraphics[scale=0.22]{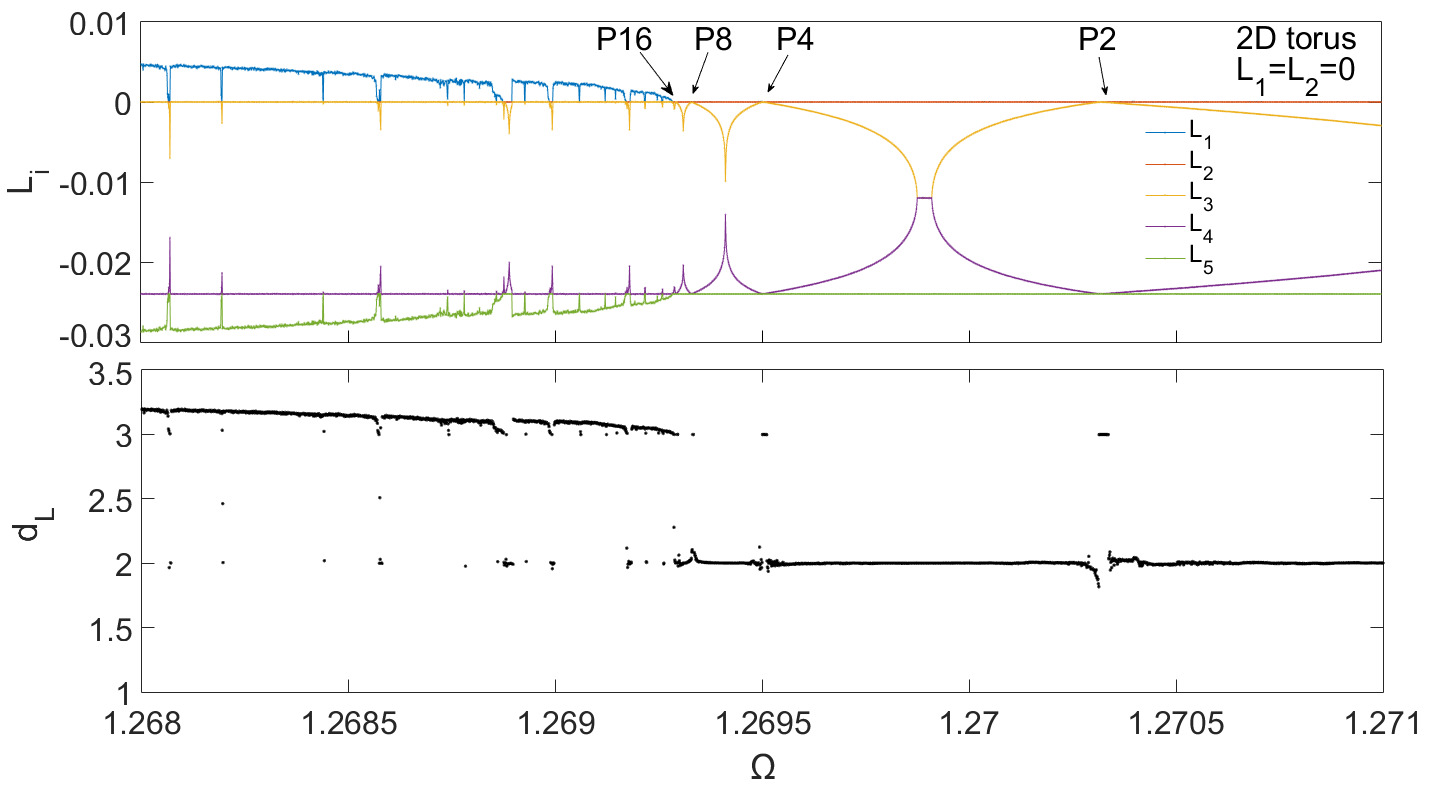}
   \caption{The complete set of Lyapunov exponents $L_i$, $i=1,...,5$ for the SQUID dimer (top) and the corresponding Lyapunov dimension (bottom), as a function of the driving frequency $\Omega$. The 
            bifurcation analysis depicts a transition from a stable 
            two-dimensional torus to chaos via a sequence of four consecutive 
            period-doubling bifurcations marked by $P_i$ ($i=2, 4, 8, 16$), where 
            the period of the solution doubles. 
            Other parameters: 
            $\lambda=0.15$, $\phi_{ac}=0.02$, $\gamma=0.024$ and $\beta=0.1369$.
            }
\label{fig7}
\end{figure}

We examine this route to chaos through period doubling of a 2D torus in our system
by fixing the magnetic coupling strength to a certain value $\lambda=0.15$ 
(marked by a black horizontal line in Fig.~\ref{fig5}(b)) and by varying the 
driving frequency $\Omega$ in a narrow interval that encloses chaotic states. 

For this choice of parameters, the resonance curve (Fig.~\ref{fig6}(a)) exhibits the same sequence of bifurcations (PB and consequently NS) as in Fig.~\ref{fig3}(a), but presents no second chaotic band and therefore no ``bubble'' connected to it. The transition into chaos is shown in Fig.~\ref{fig6}(b) where the maximum values of the magnetic flux $\phi_1$ are plotted for the same $\Omega$ interval.  

For each value of $\Omega$ the whole spectrum of the Lyapunov exponents is 
calculated and plotted in Fig.~\ref{fig7} (top). Starting from the right end of the 
frequency interval, it is observed that the two largest Lyapunov exponents are 
zero ($L_1=L_2=0$), while the other three are less than zero. This indicates 
quasiperiodic dynamics on a stable 2D torus.
When the 
frequency $\Omega$ decreases, one more Lyapunov exponent reaches zero at 
$\Omega=1.27032$ (marked as P2 on the figure), indicating a period doubling 
bifurcation of the torus. By further decreasing $\Omega$, we find three more 
points in which the three largest Lyapunov exponents are zero at 
$\Omega=1.269505$, $1.26933$, and $1.269295$ indicating three more period 
doupling bifurcations of the 2D torus (marked by P4, P8, and P16, respectively, 
on the figure). For frequencies even lower than that at P16 the 2D torus 
break down and the system enters into a chaotic state, as it is indicated by the 
positivity of the largest Lyapunov exponent ($L_1 =L_{max} >0$). 
This transition to chaos is also reflected in the 
Lyapunov dimension $d_L$, which according to Kaplan and York~\cite{KAP79} is given by:
\begin{equation}
\label{01}
  d_L = k +\frac{1}{|L_{k+1}|} \sum_{i=1}^k L_i,
\end{equation}
where $k$ is defined by the condition that
\begin{equation}
\label{02}
  \sum_{i=1}^k L_i \geq 0 ~~~{\rm and} ~~~ \sum_{i=1}^{k+1} L_i < 0.
\end{equation}
As the 2D torus transitions into chaos, the Lyapunov dimension changes from the value $d_L=2$ to $d_L>3$, as shown in Fig.~\ref{fig7} (bottom).

\begin{figure*}[t!]
   \includegraphics[scale=0.45]{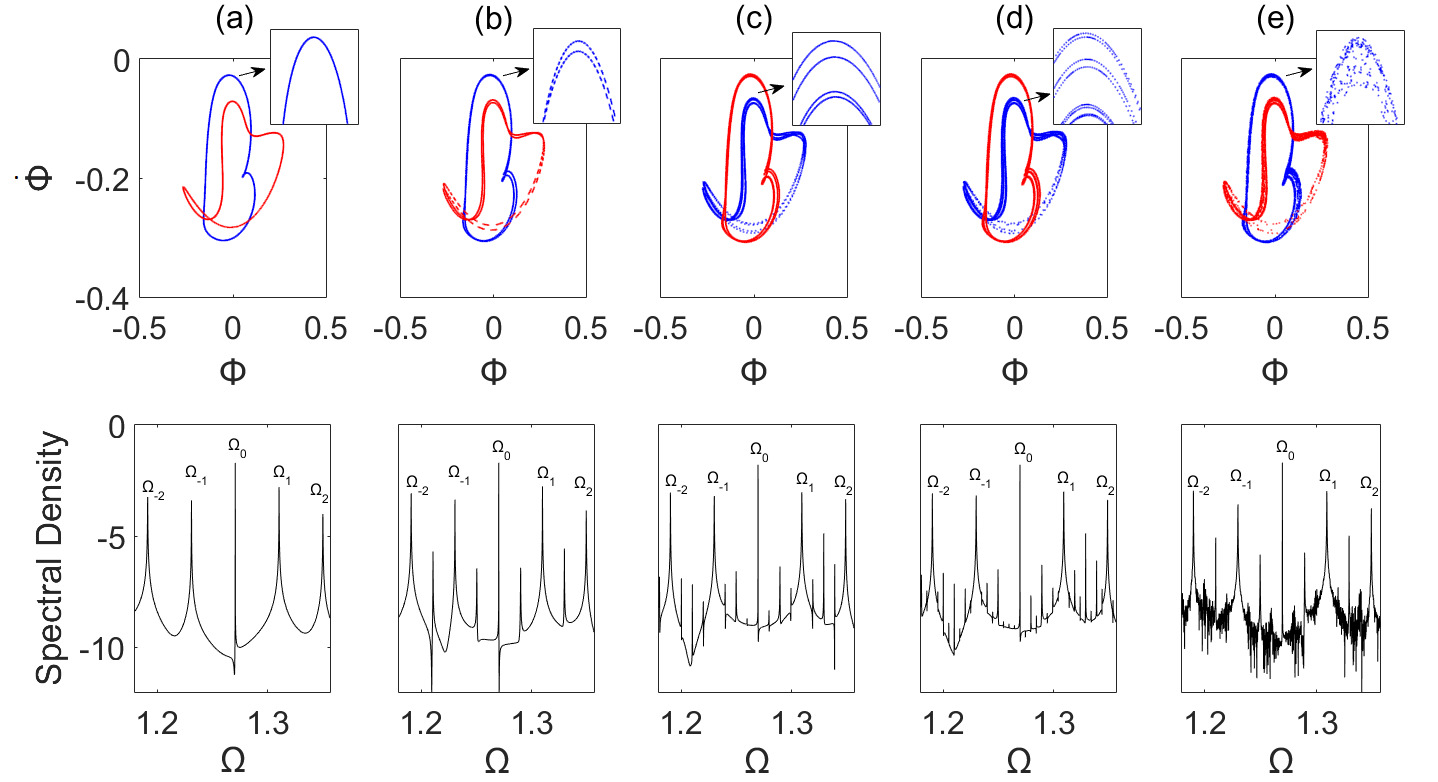}
   \caption{Poincar{\'e} cross sections on the $\Phi-\dot{\Phi}$ plane (upper 
            panels) and the corresponding frequency spectra plotted in 
            a logarithmic scale (bottom panels), for (a) $\Omega=1.2705$; 
            (b) $\Omega=1.27$; (c) $\Omega=1.2694$;(d) $\Omega=1.26931$; 
            (e) $\Omega=1.26923$. In the upper panels, blue and red lines 
            correspond to the first ($\Phi =\phi_1$, $\dot{\Phi} =\dot{\phi}_1$) 
            and second ($\Phi =\phi_2$, $\dot{\Phi} =\dot{\phi}_2$) SQUID, 
            respectively. 
            {\em Insets:} Enlargement of a small part of Poincar{\'e} sections
            of the trajectories of the first SQUID to make clear the period of 
            the tori. Other parameters as in Fig. \ref{fig7}. 
\label{fig8}}	    
\end{figure*}

Typical Poincar{\'e} sections of the period-2, -4, -8, and chaotic 2D tori 
on the $\phi_{1,2} - \dot{\phi}_{1,2}$ phase plane are shown in the upper panels 
of Figs.~\ref{fig8}(a) - (e). Blue and red lines correspond to flux threading 
the loop of the first ($1$) and the second ($2$) SQUID, respectively. The insets 
show enlargements of a small part of the Poincar{\'e} sections of the first 
SQUID (SQUID number $1$), in order to see clearly the varying number of layers 
of the tori as the frequency $\Omega$ decreases. 
The Poincar{\'e} section of the period-16 torus is very similar to that of the 
period-8 torus and thus it has been omitted. In the lower panels of 
Fig.~\ref{fig8}, the corresponding Fourier power spectral densities in a 
logarithmic scale are shown. 
The 2D torus (upper panel of Fig.~\ref{fig8}(a)) has a dominant frequency which 
in the present case coincides with the external (driving) frequency $\Omega$, 
as well as four more frequencies denoted as $\Omega_{\pm 1}$, $\Omega_{\pm 2}$
(lower panel of Fig.~\ref{fig8}(a)). The exact values of these frequencies are
obtained from the locations of the sharp peaks. Then, as the period of the 2D 
torus increases to $2$, $4$, and $8$ through period-doubling bifurcations, a 
doubling of the sharp peaks occurs (as can be seen in Fig.~\ref{fig8}(b), (c),
and (d), respectively) that correspond to frequencies with significant spectral 
content. In Fig.~\ref{fig8}(e), in which the SQUID dimer is in a chaotic state,
the power spectrum exhibits a substantial noisy background, along with several 
sharp peaks that correspond to frequencies with significant spectral content. 
Note that the driving frequency $\Omega$ is dominant in all five subfigures.

Thus, we have identified a novel route to chaos for the SQUID dimer, which 
initiates from a single limit cycle. The latter then bifurcates into a 2D
torus, which follows a finite period doubling cascade which eventually leads to 
a chaotic state. Note that in the case of two coupled Duffing oscillators with
dissipation and driving force, three period-doubling bifurcations of a 
three-torus were identified for the transition to chaos through quasiperiodicity
\cite{Kozlowski1995}.

\section{Localized $\&$ Chaotic Synchronization}
The SQUID dimer considered here consists of two symmetrically coupled identical 
nonlinear oscillators. Such systems were once believed to support only totally 
synchronous or asynchronous states. However, as it is known today, there are more
possible forms of synchronization such as the so called localized 
synchronization. This form of synchronization has been originally investigated 
in a slightly dissimilar pair of solid state \cite{Kuske1997} and semiconductor 
\cite{Hohl1997} lasers. For two coupled oscillators, localized synchronization 
means that one of the oscillators exhibits strong oscillations while the other 
one exhibits weak oscillations. In the other limit, that of asynchronous dynamics,
extreme case of having the one oscillator exhibiting periodic motion and the other 
chaotic motion has been recently reported \cite{Awal2019}. We explore below the 
effect of localized synchronization for the SQUID dimer using the correlation 
function between the time-dependent fluxes through the loop of the first and the 
second SQUID $\phi_{1}(t)$ and $\phi_{2}(t)$, respectively, i.e., 
\begin{equation} 
\label{999}
   C_{12} =\frac{\langle (\phi_{1}(t)-\mu_{1})(\phi_{2}(t)-\mu_{2}) \rangle}
                {\sigma_{1} \sigma_{2}}, 
\end{equation}
where $\mu_{1,2}$ and $\sigma_{1,2}$ are the mean value and the standard 
deviation of the corresponding time series of $\phi_1$ and $\phi_2$, 
respectively. Using this measure we can quantify the asymmetry in the amplitudes 
of the two coupled SQUID magnetic fluxes. If $\phi_{1}$ and $\phi_{2}$ are 
increasing or decreasing simultaneously then the correlation function will be 
positive. In contrast, if the value of $\phi_{1}$ is increasing while $\phi_{2}$ 
is decreasing and vice versa, then $C_{12}$ is negative. The maximum/minimum 
value of the correlation function is $\pm 1$. For $C_{12}\simeq 1$ the time 
series of $\phi_1$ and $\phi_2$ are almost identical thereby the two SQUIDs 
exhibit in-phase synchronization, whereas for $C_{12}\simeq -1$ they are in 
anti-phase synchronization.

\begin{figure*}[!t]
   \includegraphics[scale=0.4]{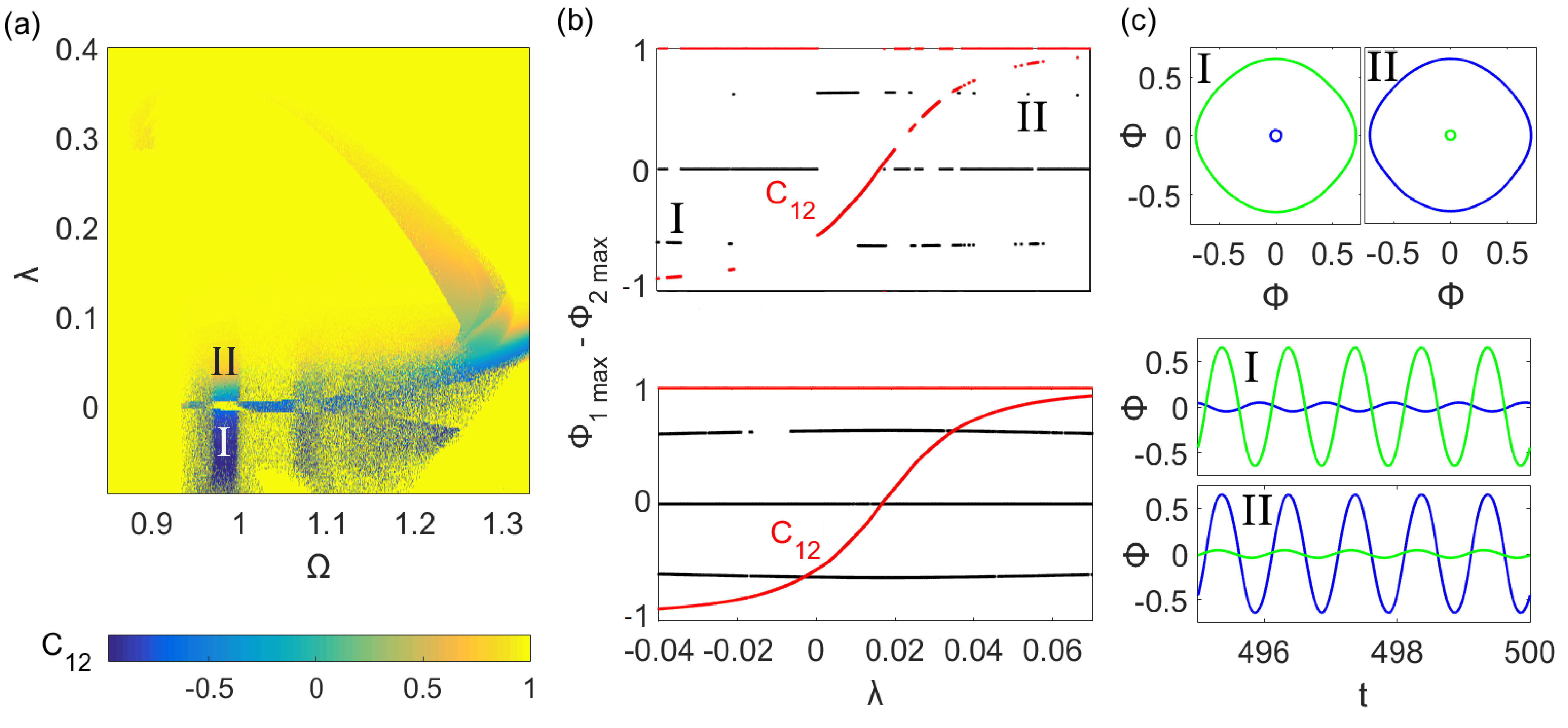}
   \caption{(a) Dynamical behavior in the ($\Omega, \lambda$) parametric space 
            of the correlation function $C_{12}$. For a completely in-phase 
            synchronized state $C_{12} =1$ while for a completely anti-phase 
            synchronized state $C_{12} =−1$. 
            (b) The correlation function (red line) and maxima differences 
            between $\phi_{1}$ and $\phi_{2}$ (black line), versus the coupling 
            strength ($\lambda$), for one random set of initial conditions (top) 
            and for $n_i =30$ different initial conditions (bottom) where 
            $\Omega=0.9822$. 
            (c) Phase portraits (top) and time series (bottom) of the magnetic 
            flux for the first SQUID, $\phi_1$ (green line), and the second 
            SQUID, $\phi_2$ (blue line), for (I) $\lambda=-0.0398$, 
            $C_{12}=-0.9103$; (II) $\lambda=0.0672$, $C_{12}=0.9248$. 
            Other parameters 
            $\phi_{ac} =0.02$, $\gamma =0.024$, and $\beta =0.1369$.} 
\label{fig9}
\end{figure*}

In Fig.~\ref{fig9}(a), the asymptotic value of the correlation function $C_{12}$  
is mapped onto the $\Omega, \lambda$ parameter space. As mentioned earlier, the 
value of $C_{12}$ reflects the dynamical behavior of the SQUID dimer as long as 
its synchronization properties are concerned. As it can be observed, the value 
of $C_{12}$ is equal to $1$ in a large area (yellow) of the parameter space but 
there also exist regions where $C_{12} \neq 1$. For example, a dark-blue region
(\Romannum{1}) can be observed in which $C_{12}$ approaches $-1$, indicating 
anti-synchronized (i.e., phase difference $\pi$ between $\phi_1 (t)$ and 
$\phi_2 (t)$) dynamics. Moreover, we observe a ``horn''-shaped region of 
intermediate to high values of $C_{12}$ which coincides with the corresponding 
``horn''-shaped chaotic region of Fig.~\ref{fig6}(b). This particular region, 
which is bounded by the Neimark-Sacker (NS) bifurcation lines, intermittent 
chaotic synchronization \cite{Baker1998,Zhao2005} is observed. In the 
case of the SQUID dimer, the effect of intermittent chaotic synchronization is
manifested by the entrainment of the two SQUID time series for $\phi_1(t)$ and 
$\phi_2(t)$ in random temporal intervals of finite duration~\cite{Zhao2005}.
This will be further investigated in future works involving 
SQUID trimers where more interesting synchronization phenomena can be observed, 
like small chimeras~\cite{Banerjee2018}.

Next we focus on the behavior of the correlation function $C_{12}$ in the 
non-chaotic region of the $(\Omega,\lambda)$ parameter space and reveal another 
interesting behavior. For a fixed $\Omega$ we study the dependence of $C_{12}$ 
as the coupling strength $\lambda$ varies from region \Romannum{1} to 
\Romannum{2} of Fig.~\ref{fig9}(a). A cross section of $C_{12}$ for 
$\Omega =0.9822$ is plotted in Fig.~\ref{fig9}(b) in red color, for one random 
set of initial conditions (top) and $n_i =30$ sets of different, randomly chosen 
initial conditions (bottom). The difference between the maximum values of 
the two SQUID magnetic fluxes $\phi_{1\text{max}}-\phi_{2\text{max}}$ is also 
plotted in black color. We make the following observations: first of all, the 
fully phase-synchronized state ($C_{12}=1$) exists for all $\lambda$ in this 
interval. 
Secondly, the correlation function appears to have a (co)sinusoidal dependence 
on the coupling strength $\lambda$ as expected, since the correlation function 
of two out-of-phase cosine signals is a cosine function of their phase 
difference \cite{Oppenheim1996}. 
In the low-coupling non-chaotic regions \Romannum{1} and \Romannum{2} 
we can assume that the two SQUID solutions are also (co)sine-like. For 
$\phi_{1}(t)$ and $\phi_{2}(t+\tau)$ or vice versa for $\phi_{1}(t+\tau)$ and 
$\phi_{2}(t)$, $\tau$ is the phase difference between the SQUID fluxes and is 
dependent on the set of initial conditions. From Fig.~\ref{fig9}(b) (bottom) we 
may claim that the $\tau=0$ case corresponds to $C_{12}=1$ and the 
(co)sinusoidal-like form of $C_{12}$ corresponds to the case where $\tau$ is a 
function of $\lambda)$. In the former case the two SQUIDs are in phase and 
$\phi_{1\text{max}}-\phi_{2\text{max}}$ is zero, while in the latter case their 
phase difference varies from values close to $-1$ (almost anti-phase) to values 
close to $+1$ (almost in-phase). The difference 
$\phi_{1\text{max}}-\phi_{2\text{max}}$ attains a nonzero positive/negative 
value depending on whether $\phi_1$ is larger/smaller than $\phi_2$, for all 
$\lambda$ and all initial conditions. Therefore this case corresponds to an 
asymmetric solution as far as the amplitudes of the two coupled SQUIDs are 
concerned.

This is illustrated in Fig.~\ref{fig9}(c) in the phase portraits (top) and time 
series (bottom) of the magnetic flux for the first SQUID (green line) and the 
second SQUID (blue line) for parameters tha belong to region \Romannum{1} 
$(\lambda=-0.0398, C_{12}=-0.9103)$ and region \Romannum{2} 
$(\lambda=0.0672, C_{12}=0.9248)$. These plots show the phase portraits of the 
corresponding stable periodic solutions where large asymmetry in the amplitudes of 
the two coupled SQUIDs occurs. Due to that difference between the amplitudes, 
these two states are illustrative examples of localized synchronization. 
Moreover, through the time series, it can be clearly seen that the fluxes 
$(\phi_{1}$  and $\phi_{2})$ are almost out of phase (I) and almost in phase 
(II) because the correlation function has values close to $-1$ and $1$, 
respectively.



\section{Conclusions}
In summary, a dimer comprising two magnetically coupled SQUIDs is investigated 
numerically with respect to its bifurcation structure and its transitions from 
quasiperiodicity to chaos through a finite period-doubling cascade of tori. 
The SQUID dimer is a prototype, dissipative-driven dynamical system of 
considerably complex dynamic behavior, on which numerous nonlinear dynamic 
effects can be investigated. In the present work, the bifurcation structure of 
the SQUID dimer was obtained using the driving frequency $\Omega$ and the 
coupling strength $\lambda$ as external control parameters that form a 
two-dimensional parameter space. Bifurcation (stroboscopic) diagrams as a  
function of $\Omega$ reveal multibranched resonance curves resulting through 
different kinds of bifurcations, transitions to chaos after four period-doubling 
bifurcations of a 2D torus, and riddle-like basins. Bifurcation diagrams for 
SN and NS bifurcations are shown on the $\Omega - \lambda$ plane. In combination
with maps of the maximum Lyapunov exponent $\Lambda_{max}$, it is observed that
the areas of the $\Omega - \lambda$ plane where chaotic behavior is expected is
bounded by the NS bifurcation curves. Eventually, localized synchronization is 
observed and quantified in the SQUID dimer using the correlation function 
$C_{12}$. The possibility of intermittent chaotic synchronization is briefly 
mentioned. 
\begin{acknowledgments}
This work was supported by the General Secretariat for Research and Technology 
(GSRT) and the Hellenic Foundation for Research and Innovation (HFRI) (Code: 203). 
JS acknowledges support by the Ministry of Education and Science of the Russian 
Federation in the framework of the Increase Competitiveness Program of NUST 
``MISiS'' (Grant number K4-2018-049). The authors would like to thank 
S. M. Anlage for fruitful discussions.
\end{acknowledgments}

\end{document}